\RequirePackage{fix-cm}
\documentclass[injp]{article}
\usepackage{graphics}
\usepackage{amsmath}
\usepackage{graphicx}
\usepackage{amssymb}
\usepackage{epsfig}
\usepackage{color}
\usepackage{subfigure}
\def\etal{et al.\ }

\def\be{\begin{equation}}
\def\ee{\end{equation}}

\def\nc2{\left( \begin{array}{c} n \\ 2 \end{array}\right)}

\def\beq{\begin{equation}}
\def\eeq{\end{equation}}

\def\n{\noindent}
\def\ij{RR^{\prime}}

  \def\ket{\vert \vert  \{ \emptyset \} \rangle}
  \def\ket2{\vert \vert \otimes \{ R \} \rangle}

\def\.#1{\mathaccent 95#1}
\def\^#1{\mathaccent 94 #1}
\def\~#1{\mathaccent "7E #1}

\def\eq{\enskip =\enskip}

\def\mns{\ -\ }

\def\trans{{\cal T}}
\def\proj{{\cal P}}

  \def\proj{{\cal P}}
  \def\trans{{\cal T}}
  \def\ket{\vert \vert  \{ \emptyset \} \rangle}
  \def\ket2{\vert \vert \otimes \{ R \} \rangle}

  \def\ket{\vert \vert  \{ \emptyset \} \rangle}
  \def\ket2{\vert \vert \otimes \{ R \} \rangle}

  \def\ket{\vert \vert  \{ \emptyset \} \rangle}
  \def\ket2{\vert \vert \otimes \{ R \} \rangle}

\def\.#1{\mathaccent 95#1}
\def\^#1{\mathaccent 94 #1}
\def\~#1{\mathaccent "7E #1}

\def\eq{\enskip =\enskip}

\def\trans{{\cal T}}
\def\proj{{\cal P}}

\begin{document}
\begin{center}
{\LARGE\bf{Configuration and Self-averaging in disordered systems}}
\vskip 1cm
{\bf S. Chowdhury$^{1}$, D. Jana$^{1,\ast}$, B. Sadhukhan$^{2}$, D. Nafday$^3$, S. Baidya$^3$, T. Saha-Dasgupta$^3$ and A. Mookerjee$^{2,3,4}$}
\vskip 1cm
{\baselineskip 7pt $^1$Department of Physics, University of Calcutta, 92 Acharya Prafulla Chandra Road, Kolkata 700009,W.B., India}\\
{\baselineskip 7pt $^2$Department of Physics, Presidency University, 86/1 College Street, Kolkata 700073, W.B., India.}\\
{\baselineskip 7pt $^3$Department of Condensed Matter and Materials Science,  S.N. Bose National  Centre for Basic Sciences, JD-III, Salt Lake, Kolkata 7000098, W.B., India.}\\
{\baselineskip 7pt $^4$Department of Physics, Lady Brabourne College, 1/2 Suhrawardy Street, Kolkata 700017, W.B., India.}
\end{center}
 
\vfill \hrule
\vskip 0.1cm
\noindent $^\ast$Corresponding Author, Email : cujanad@yahoo.com\\     
             
\date{Received: date / Accepted: date}
\newpage

\begin{center}
 {\bf Abstract}
\end{center}

The main aim of this work is to present two different methodologies for configuration averaging
in disordered systems. The Recursion method is suitable for the calculation of spatial
or self-averaging, while the Augmented space formalism averages over different possible configurations of the system. We have applied these 
techniques to a simple  example and compared their results. Based on these, we have reexamined the concept of spatial ergodicity in disordered systems. The 
specific aspect, we have focused on, is the question ``Why does an experimentalist often obtain the averaged result on a single sample ?" We have found 
that in our example of disordered graphene, the two lead to the same result within the error limits of the two methods.
 \vskip 1cm


\newpage
\section{\bf Introduction : Averaging in disordered systems}

The study of averaging over all possible different  `configurations' of a quenched disordered system  has been a focused problem in the theory of 
measurements. Configuration averaging is ubiquitous  both in quantum mechanics and statistical physics. For annealed disorder, where the 
thermal disorder driven fluctuations control the behavior of the system, the idea of spatial ergodicity is important. At finite temperatures 
different possible states of a canonical ensemble, for example, are  occupied with Boltzmann probabilities, and observable physical properties 
are averaged over the ensemble. Similarly, when we wish to measure a given physical observable in a quantum system, the result of the 
measurement is spread over different possible states with probabilities given by  squared amplitudes of their wave function projection 
onto those states. Our discussions will be essentially at 0K and we shall focus on frozen or quenched disorder as in a glass or disordered 
alloy. 

During the last four decades 
considerable effort has gone into devising methods for carrying out averages of physical observables over
 different configurations realized by disordered systems.
 
  Why do we wish to carry out such averages and is such a  procedure meaningful \cite{kn:erg1}?
  
  Let us examine a specific example. An experimentalist is carrying out energy resolved photo-emission  studies on a disordered binary alloy 
  A$_x$B$_y$. Varying the frequency of the incident photon and keeping the energy window of the excited outgoing electrons reasonably 
  narrow, one can map out the density of states of the valence electrons for the alloy. Ideally, if the experiment is carried out on ten 
  different samples of the same alloy, slightly different results should be obtained. Every different sample of the disordered alloy has 
  different distributions of the A and B  constituents and hence should give slightly different random results.  Yet, in practice, the 
  variation, the experimenter sees in the different samples, is well within experimental error bars. What is obtained is an average result, 
  averaged over different realizable configurations of atomic arrangements in the alloy. The interesting fact is that averaged result is obtained in a 
  single large sample. The same is true for other measured bulk properties like the specific heat, 
  conductivity and different response functions. 

Note that  all these measured properties are  global. Should there be a difference if we measure local properties with local probes ? Let us take 
another example of a magnetically disordered alloy AuFe (with $<$ 10$\%$ of Fe). If we  measure the magnetization of a sample, it remains zero 
upto liquid He temperatures. Yet, if we carry out a M\"ossbauer study on the same alloy, there is a clear indication of a frozen local exchange 
field at low temperatures, indicating the existence of non-zero local magnetization. Configuration averaging is meaningless if we wish to 
look at local properties. Even here, a degree of averaging over the far environment is relevant. Although the radioactive Fe atom giving rise 
to the M\"ossbauer spectrum sits in different environments in different samples, yet experiments yield an average 
exchange field distribution.                                

We shall focus on the question, ``Why does an experimentalist working on a disordered material often obtain the averaged result from a single sample ?"         

\section{Understanding self-averaging}

\begin{figure}[t]     
\centering           
\includegraphics[width=10cm,height=10cm]{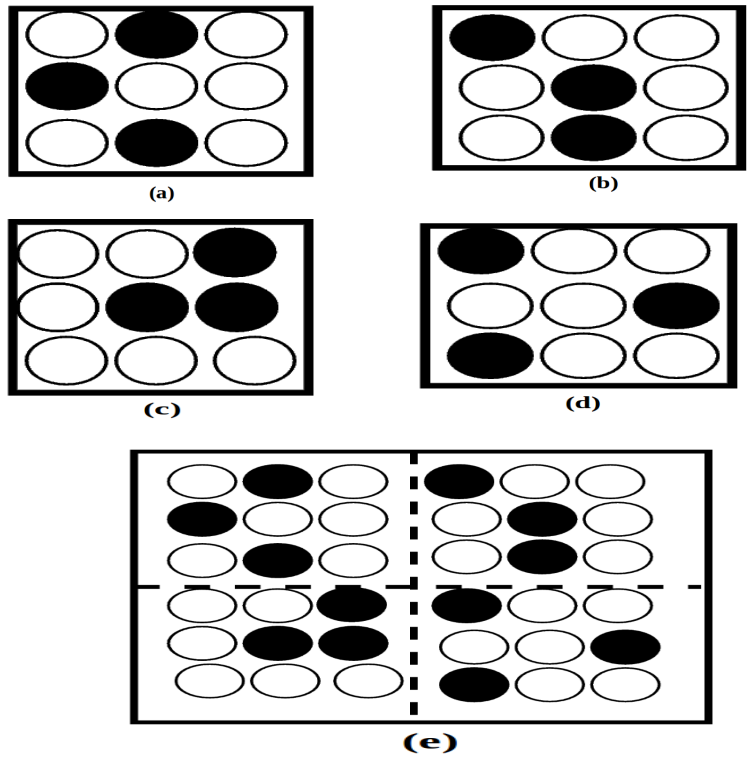}\hskip 1.0cm
\caption{(a)-(d) Ensemble of macrosystems in the  'configuration space' of a binary alloy. (e) A single macrosystem constructed out of the microsystems shown in 
(a)-(d). \label{fig1}}
\end{figure}

Why do we observe configuration averaged results in a particular macroscopic sample~? The answer lies in the idea of spatial ergodicity 
or self-averaging. We shall use these terms as synonyms. Let us look at the example of a random binary alloy as illustrated in 
Figs.\ref{fig1} (a) - \ref{fig1} (d). They show a number of different samples each with N atoms and its own A-B distribution. There are in all 2$^N$ 
distinct  configurations. Let us label each configuration by ${\cal C}_n$, formally the configuration average of a physical quantity ${\cal A}$ 
is 

\begin{equation}
 \ll {\cal A} \gg_{conf} = \frac{1}{2^N} \sum_n {\cal A}({\cal C}_n) \gamma({\cal C}_n)
\end{equation}

where $\gamma({\cal C}_n)$ is the number of times a given configuration ${\cal C}_n$ occurs in the collection (space) of configurations shown 
in Fig.\ref{fig1}(a)-(d). Now, if $N\rightarrow\infty$ then $\gamma({\cal C}_n)/2^N$ is the probability associated with the configuration and 
\begin{equation} 
 \ll {\cal A}\gg_{conf} = \sum_n {\cal A}({\cal C}_n) {\cal P}r({\cal C}_n)
\end{equation} 

Let us now take a large single sample with 2$^N$ atoms and partition it into subsystems each with $N$ atoms. The spatial average taken over 
this one single sample is :

\begin{equation}
 \ll {\cal A}\gg_{spat} = (1/2^N)\sum^{2^N}_{n=1} {\cal A}(\vec{r}_n)
\end{equation} 

\noindent where $\vec{r}$ denotes the positions of the atoms. We shall now partition the sample
as shown in Fig.\ref{fig1} (e) and group these sites into microsystems of size $N$ atoms, so there are $2^N/N$  such partitions
which we shall call ${\cal C}'_n$, then

\begin{equation}
 \ll {\cal A}\gg_{spat} = (N/2^N) \sum_{m=1}^{2^N/N} \frac{1}{N} \sum_{m\in {\cal C}'_m} {\cal A}(\vec{r}_m)
\end{equation}
 In the collection ${\cal C}'_m$ all distributions are not distinct. Assume that there are $\gamma({\cal C}'_m)$ identically distributed 
 microsystems, then the above equation can be written as :
 
 \begin{equation}
 \ll {\cal A}\gg_{spat} = (N/2^N) \sum_{m=1} {\cal A}({\cal C}'_m) \gamma({\cal C}'_m)  
 \end{equation}
 
If we now we let $N \rightarrow\infty$ then we get :
 \begin{equation}
 \ll {\cal A}\gg_{spat} =  \sum_{m=1} {\cal A}({\cal C}'_m) {\cal P}r({\cal C}'_m)  
 \end{equation}

Is it then true that it follows from Eqs. (1) and (5) that the configuration average taken over many different samples is the same as the 
spatial average over a single sample ? The answer is in general in the negative. We should note that :

\begin{itemize}
\item [(i)] The statement is untrue for any finite system.
\item [(ii)] The statement remains true if, as $N\rightarrow\infty$ in such a way that each partition of the single sample also becomes 
infinitely large, but for every configuration (shown on Fig.\ref{fig1} (a)-\ref{fig1} (d)) there is a one-to-one correspondence with a partition shown in 
Fig.\ref{fig1} (e) and vice versa. This is a very strong statement and is known as the "Spatial Ergodic Principle" for quenched 
disordered systems. \item[(iii)] If these averages diverge as $N\rightarrow\infty$, but the variance diverges faster, then there is no point in 
talking about averages, since the fluctuations about the average dominate. Example of such a system is the intensity of starlight after it 
passes through a disordered dielectric medium. The fluctuations in intensity dominate over the average, causing star to twinkle even outside 
our atmosphere. \cite{kn:cs}. The same holds for conductance fluctuations in disordered media.\cite{mj}.
\end{itemize}

Although numerous very detailed and rigorous works exist on temporal ergodicity \cite{kn:erg7,kn:erg5}, a similar detailed exposition on 
spatial ergodicity is scarce. The concept of spatial ergodicity is a conjecture; its mathematical proof involves many stringent pre-conditions. 
Many systems do not satisfy them and therefore are not spatially ergodic. In order to develop an algorithm for configuration averaging, 
we have to be careful to ensure that the assumption of spatial ergodicity remains valid. The aim of this paper is to introduce two different 
numerical techniques : one of which explicitly calculates the spatial average and the other the configuration average and then compare the two 
results for a simple example of disordered graphene.  We have specifically chosen a two-dimensional model, as prior experience tells us that non-ergodicity appears more often in lower dimensional systems \cite{mj}.

\section{\bf Spatial averaging and the Recursion Method }

Our systems of interest are disordered systems, and consequently the Bloch Theorem fails to hold. 
Strictly speaking reciprocal space approaches are invalid. There has been supercell approaches, where the local disorder has been incorporated 
in a large supercell and periodic boundary conditions were imposed on the surfaces of the supercell. Unfortunately, the fact that whenever 
there are imposed periodicities, the spectral function always leads to sharp bands and the disorder induced lifetimes which gives a spread to 
the bands cannot be obtained. In all such approaches one has to introduce a small imaginary part to the energy to smoothen these sharp features 
and this is introduced arbitrarily `by hand'. We have briefly discussed the effects of the superlattice approach and its attendant artificial 
periodicity imposed in 
Appendix 1.

We turn instead to real space techniques. We first propose a methodology to deal with spatial averaging in disordered systems.  
It has been argued that many of the properties of solids are crucially dependent on the local chemistry of the atoms 
constituting the solid \cite{ter2}. For such properties a Black Body Theorem essentially states that the very far environment of 
an atom in a solid has very little influence on its local chemistry. This local environment approach to the electronic structure of solids
requires an alternative to band theory for solving the Schr\"{o}dinger equation. Band theory is invalid in disordered systems.  Physics is 
better understood by means of a solution that explicitly accounts for the role of local environment. The recursion method introduced by 
Haydock {\it et al}\cite{md27} is a lucid approach in this direction. It expresses the Hamiltonian in a form that couples an atom to its first 
nearest neighbor, then through them to its distant neighbors and so on. Mathematically,  a new orthonormal basis set $\vert n \rangle$ is 
constructed  by a three term recurrence formula to make the Hamiltonian tridiagonal.The starting state $\vert 0 \rangle$ of recursion is  :
\be \vert 0\rangle = \frac{1}{\sqrt{N}}\sum_i \ \eta_i\vert R_i\rangle \ee

where, $\eta_i$ take the values $\pm 1$ randomly, so that :

\begin{eqnarray*}\langle 0\vert G(z)\vert 0\rangle &
  = \frac{\displaystyle 1}{\displaystyle N} \left\{ \sum_i \eta_i^2 \langle R_i\vert G(z)\vert R_i\rangle +
                            \sum_i\sum_j \eta_i\eta_j\langle R_i\vert G(z)\vert R_j\rangle \right\}\end{eqnarray*}

Since $\eta_i^2=1$ and the second term ~$O(\sqrt{N})$, the spatial average of the total density of states (TDOS) is :
\be -\frac{1}{\pi}\ \Im m\ Tr G(E+i0) = \ll n(E)\gg_{\rm sp} \ee                  

The whole set of orthonormal states are generated by the following three
term recurrence relation:
\begin{equation}
\beta_{n+1}\vert n+1 \rangle = H\vert n \rangle - \alpha_{n} \vert n \rangle - \beta_{n} \vert n-1 \rangle
\label{rec1}\end{equation}
 
Since we cannot apply Bloch theorem for systems where periodic symmetry is lost, we take recourse to an alternative approach of obtaining 
physical properties  from the averaged resolvent. Haydock \etal \cite{md27} have showed that using Eq. (\ref{rec1}) we can expand the
resolvent as a continued fraction :

\be
 \ll G_{RR}(z)\gg_{\rm sp} = 
 \frac{1}{\displaystyle z-\alpha_{0}
            -\frac{\beta_{1}^{2}}{\displaystyle z-\alpha_{1}
            -\frac{\beta_{2}^{2}}{\displaystyle z-\alpha_{2}
            -\frac{\beta_{3}^{2}}{\displaystyle z-\alpha_{3}
            -\frac{\beta_{4}^{2}}{\displaystyle z - \alpha_4 - T(z)}}}}}             
\hfill\label{cf}
\ee

In practice, the continued fraction is evaluated to a finite number of steps.

\subsection{\bf The far environment : terminators}

Right at the start we chose the real space algorithm over mean-field and supercell approaches because we do not wish to introduce artificial 
periodicity and miss out on the effects of long-ranged disorder. The problem with any numerical calculation is that we can deal with only a 
finite number of operations. In the recursion algorithm, we can go up to a finite number of steps and if we stop the recursion, or impose 
periodicity this would lead to exactly what we wish to avoid. The analysis of the asymptotic part of the continued fraction is therefore of 
prime interest to us. This is the "termination" procedure discussed by Haydock and Nex\cite{hn}, Luchini and Nex\cite{ter3}, Beer and 
Pettifor\cite{bp} and in considerable detail by Viswanath and Muller \cite{ter5}. This terminator $T(z)$ which accurately describes the far 
environment, must maintain the herglotz analytical properties. We have to incorporate not only the singularities at the band edges, but also 
those lying on the compact spectrum of $H$. Viswanath and Muller \cite{ter5} have proposed a terminator :

\be   T(z) = \frac{2\pi(E_m)^{(p+2q+1)/2}}{B\left(\displaystyle{\frac{p+1}{2}},1+q\right)} \vert z-E_0\vert^p\left\{(z-E_1)(E_2-z)\right\}^q \ee

The spectral bounds are at $E_1$ and $E_2$ with square-root singularities, $E_m^2 = E_1E_2$ and there is a cusp singularity at $E_0$ if $p=1,q=1$ or 
infra-red divergence if $p=-1/2,q=0$. $E_0$ sits on the compact spectrum of $H$. Magnus \cite{mag} has cited a closed form of the convergent 
continued fraction coefficients of the terminator :

\begin{eqnarray}
\beta^2_{2n} & = & E_m^2\ \frac{4n(n+q)}{(4n+2q+p-1)(4n+2q+p+1)} \nonumber\\
\beta^2_{2n+1} & = & E_m^2\ \frac{(2n+2p+1)(2n+2q+p+1)}{(4n+2q+p+1)(4n+2q+p+3)}\nonumber\\
\end{eqnarray}

\begin{figure}
\begin{center}
\includegraphics[height=12cm,width=12cm]{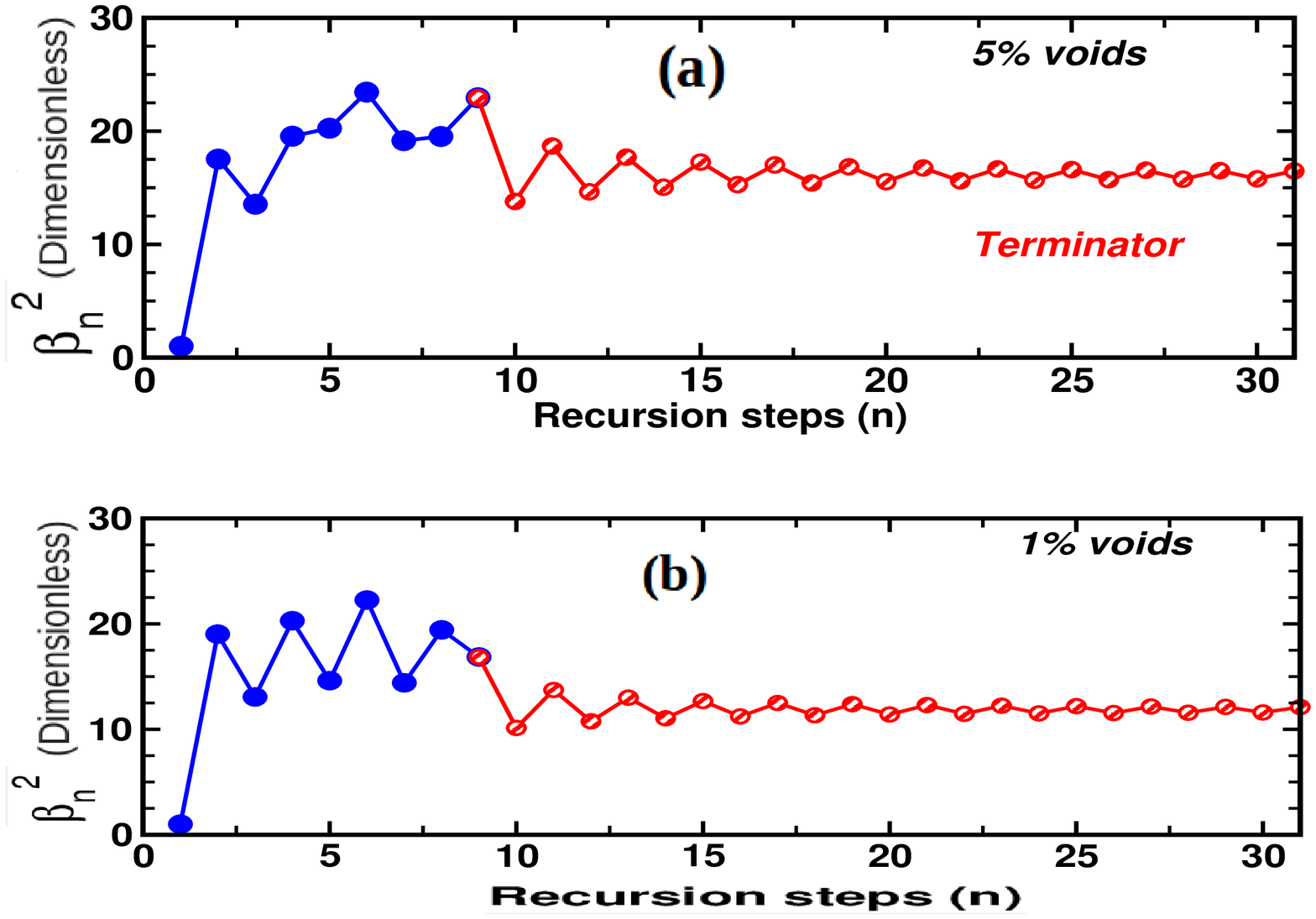}\vskip 1.2cm
\end{center} 
\caption{(a) Asymptotic parts of the calculated graphene Green function continued fraction coefficients obtained by recursion for $5 \%$ voids and  
(b) the same for $1 \%$ voids. \label{fig2}}
\end{figure}

The parameters of the terminator are estimated from the asymptotic part of the continued fraction coefficients calculated from our recursion.
Viswanath-M\"uller termination was used and seamlessly enmeshed with the calculated coefficients as shown in Fig. \ref{fig2} (a) and \ref{fig2} (b). In 
this way both the near and the far environments are accurately taken into account.

\section{\bf Ensemble averaging and the Augmented Space technique}

The augmented space or configuration space method had been proposed as early as 1973 by Mookerjee \cite{mook}. Some of the most successful 
beyond single-site, mean-field averaging techniques in random systems are based on this method : e.g. the travelling Cluster Approximation 
(CA) \cite{tca} and the itinerant Coherent Potential Approximation (CPA) \cite{icpa}. In this approach we work not only with individual 
configurations of the sample but also with collection of all possible configurations. Let us examine the basic concepts in this methodology.

Suppose we measure a property of the system which is not disordered. In that case we may
carry out repeated measurements of the property N and get a set of results which are all the
same within error bars :  $n \pm \delta$ and we get this with probability one. When the system is disordered the same    
set of measurements yield a set of values  $n_1,n_2 \ldots n_m \pm \delta $ with probabilities $p_1,p_2\ldots p_m$. We cannot associate a 
scalar quantity with this property.
Instead, we associate an operator  $\tilde{N}$ whose eigenvalues are the measured quantities and whose
spectral density is the probability density of $\tilde{N}$.

Take for example, a random variable $n_R$, which takes the value 1 if the site $R$ is occupied
by a A type atom with probability x and 0 if the site $R$ is occupied by a B type  atom with
probability y. The next problem is the inverse problem of recursion. There the tri-diagonal representation of the Hamiltonian is given and the 
resolvent is obtained as a continued fraction. Here, the resolvent is given and by inspection, related to the probability density :

\begin{eqnarray}
 P(n_R) & = & x \delta(n_R-1) + y \delta(n_R) = - (1/\pi) Im \left\{\frac{x}{n_R-1}+\frac{y}{n_R}\right\}\\
        & = & - (1/\pi) Im   \dfrac{1}{n_R - x - \dfrac{xy}{n_R - y}} \end{eqnarray}
        
        So  \[ \tilde{N} = \left( \begin{matrix}
                       x & \sqrt{xy}   \\
                       \sqrt{xy}& y
\end{matrix}         \right) \]
          
The eigenvalues are 0 and 1 as expected. The above representation is in a basis :\\
 ${\framebox{\color{white}{\rule{0.06cm}{0.06cm}}}}_R\ =\ \sqrt{x}|0\rangle + \sqrt{y} |1\rangle$ \ ;\ 
  ${\color{black}{\rule{0.3cm}{0.3cm}}_R}\ =\ \sqrt{y}|0\rangle - \sqrt{x} |1\rangle$ \\

Each member of the basis, is a
`configuration' of $\tilde{N}$ at $R$. For later convenience we choose the first to
be the `reference configuration' and the second to be  a disorder induced `configuration fluctuation' whose creation from the reference 
configuration is described by $b^\dagger_{R}$. Since the disorder is binary and each site can have only one fluctuation, these fluctuations 
behave like fermions :
 
\[ b^\dagger_{R}\ {\color{black}{\rule{0.3cm}{0.3cm}}}_R = 0\quad ;\quad  b_{R}\ \framebox{{\color{white}{\rule{0.1cm}{0.1cm}}}}_R = 0 \]

Each site $R$ has an operator and a configuration space associated with it. The system of N sites then has a configuration space $\Phi = \prod^\otimes \phi_{R}$  of rank 2$^N$. The number of fluctuations in a configuration is called its 
`cardinality'  and the sequence of sites where they occur is called its 
`cardinality sequence'. Some typical configurations are shown below with their cardinalities
and cardinality sequences :

\begin{eqnarray*}
\quad\left\{    \framebox{{\color{white}{\rule{0.15cm}{0.15cm}}}}_{R_1}
\quad\framebox{{\color{white}{\rule{0.15cm}{0.15cm}}}}_{R_2}
\quad\framebox{{\color{white}{\rule{0.15cm}{0.15cm}}}}_{R_3}
\quad\framebox{{\color{white}{\rule{0.15cm}{0.15cm}}}}_{R_4}
\quad\framebox{{\color{white}{\rule{0.15cm}{0.15cm}}}}_{R_5}
\quad\framebox{{\color{white}{\rule{0.15cm}{0.15cm}}}}_{R_6}\quad \ldots\right\} & \mbox{card} = 0,& \mbox{card. sequence} = \{\emptyset\}\\
\left\{     {\color{black}{\rule{0.5cm}{0.5cm}}}_{R_1}
\quad\framebox{{\color{white}{\rule{0.15cm}{0.15cm}}}}_{R_2}
\quad{\color{black}{\rule{0.5cm}{0.5cm}}}_{R_3}
\quad\framebox{{\color{white}{\rule{0.15cm}{0.15cm}}}}_{R_4}
\quad\framebox{{\color{white}{\rule{0.15cm}{0.15cm}}}}_{R_5}
\quad{\color{black}{\rule{0.5cm}{0.5cm}}}_{R_6} \ldots\right\}
 & \mbox{card} = 3,& \mbox{card. sequence} = \{R_1,R_3,R_6\} \\
 \left\{    \framebox{{\color{white}{\rule{0.15cm}{0.15cm}}}}_{R_1}
\quad\framebox{{\color{white}{\rule{0.15cm}{0.15cm}}}}_{R_2}
\quad{\color{black}{\rule{0.5cm}{0.5cm}}}_{R_3}
\quad{\color{black}{\rule{0.5cm}{0.5cm}}}_{R_4}
\quad\framebox{{\color{white}{\rule{0.15cm}{0.15cm}}}}_{R_5}
\quad\framebox{{\color{white}{\rule{0.15cm}{0.15cm}}}}_{R_6}\ldots\right\} &\mbox{card} = 2,& \mbox{card. sequence} = \{R_3,R_4\}
\end{eqnarray*}

These operators are in the `configuration' space $\Phi$ of rank 2$^N$.
They may be written as :

\[ \tilde{N}_{R} =  I + (y-x) b^\dagger_{R} b_{R} + \sqrt{xy} ( b^\dagger_{R} + b_{R}) \] 

The next thrust in this approach came with the Augmented Space Theorem stated by Mookerjee \cite{mook}. We shall try to understand this by an 
example :

\be
P(n_R)  \eq (-1/\pi)\; \Im m\; \langle \emptyset \vert \Big ( (n_R  +\imath 0)I \mns  N_{R}\Big )^{-1} \vert \emptyset \rangle \eq (-1/\pi)\; \Im m\; g(z)\ee

where $g(z) = \langle\emptyset\vert (z\tilde{I}-\tilde{N}_R)^{-1} \vert\emptyset\rangle$ is the resolvent of $\tilde{N}_R$ \\

and

\[
 |\emptyset\rangle  \eq \left\vert \framebox{\color{white}{\rule{0.15cm}{0.15cm}}}_{R_1}\quad\framebox{\color{white}{\rule{0.15cm}{0.15cm}}}_{R_2}\quad\framebox{\color{white}{\rule{0.15cm}{0.15cm}}}_{R_3}\quad\framebox{\color{white}{\rule{0.15cm}{0.15cm}}}_{R_4}\quad\framebox{\color{white}{\rule{0.15cm}{0.15cm}}}_{R_5}\quad\framebox{\color{white}{\rule{0.15cm}{0.15cm}}}_{R_6}\quad \ldots\right.\rangle 
\]

Let $f(n_1,n_2 \ldots n_k \ldots)$ be a well behaved function of the
set of independent random variables $\{n_1,n_2,\ldots n_k \ldots\}$, then

\begin{eqnarray*}
  \ll f(n_1,n_2 \ldots ) \gg & = & \int dn_1 \int dn_2 \ldots 
 f(n_1,n_2 \ldots ) P(n_1)P(n_2)\ldots  \\
& = & \oint dz_1 \oint dz_2 \ldots f(z_1,z_2 \ldots )\ g(z_1) g(z_2)  \ldots
\end{eqnarray*}
Using the completeness of the eigenbasis of $\{\tilde{N}_R\}$
\begin{eqnarray*}
 \langle\emptyset| \oint dz_1 \oint dz_2\ldots f(z_1,z_2,\ldots)\int d\rho(\mu_1) \int d\rho(\mu_2) \ldots |\mu_1,\mu_2 \ldots\rangle\langle \mu_1,\mu_2 \ldots | \ldots\\
\prod (zI-N_R)^{-1}\int d\rho(\mu'_1)\int d\rho(\mu'_2)\ldots |\mu'_1,\mu'_2 \ldots\rangle\langle \mu'_1,\mu'_2 \ldots |\emptyset\rangle \\
\end{eqnarray*}

Now using the orthogonality of the eigenstates of $\{\tilde{N}_R\}$ :

\[= \langle\emptyset| \int d\rho(\mu_1) \int d\rho(\mu_2) \ldots
|\mu_1,\mu_2 \ldots\rangle\left[ \oint dz_1 \oint dz_2 f(z_1,z_2,\ldots) \prod (z-\mu_k)^{-1} \right] \langle \mu_1,\mu_2 \ldots |\emptyset\rangle
\]
\[= \langle\emptyset| \left[ \int d\rho(\mu_1) \int d\rho(\mu_2) \ldots
|\mu_1,\mu_2 \ldots\rangle  f(\mu_1,\mu_2,\ldots) \langle\mu_1,\mu_2 \ldots \right] |\emptyset\rangle
\]
Finally,
\[ \ll f(n_1,n_2 \ldots ) \gg = \langle\emptyset| \widetilde{f}(\tilde{N}_1,\tilde{N}_2 \ldots) |\emptyset\rangle \]

The operator on the right-hand side is the same operator function of the operators $N_1,N_2,\ldots$ as the function f was of $n_1,n_2 ,\ldots n_3$.
This is the central result of the augmented space technique : configuration averages are 
specific matrix elements in the full augmented space, which carries not only the information
about the underlying lattice, but also how that lattice is randomly decorated with the two
constituents subject to their concentrations.

Let us return to the  disordered binary alloy described by the Hamiltonian :

\begin{eqnarray}
 H&\ =\ &\sum_{R}\ \left[ \epsilon_B + (\epsilon_B-\epsilon_A) n_{R}\right]\ a^\dagger_{R} a_{R} \ +\ldots\nonumber\\
 &\ldots + &  \sum_{R}\sum_{\chi} \ \left[t^{BB}(\chi) + t^{(2)}(\chi)\ n_{R}\ n_{R+\chi} + t^{(1)}(\chi)\ (n_{R}+n_{R+\chi}) \right]\  a^\dagger_{R+\chi} a_{R} \end{eqnarray}
 
 Here, $t^{(1)}(\chi) = t^{AB}(\chi)-t^{BB}(\chi)$  and  $t^{(2)}(\chi) = t^{AA}(\chi)+t^{BB}(\chi)-2 t^{AB}(\chi)$,  where $\chi = R-R'$.

This Hamiltonian is an operator in a Hilbert space ${\cal H}$ spanned by the denumerable basis
$\{|R\rangle\}$. We now turn to the augmented space formalism and associate with each random
variable $n_{R}$ an operator $\tilde{N}_{R}$ such that its eigenvalues are the values attained by $n_{R}$ and
the density of states of $N_{R}$ is the probability density of $n_{R}$ for attaining its values. For
binary randomness (e.g. $n_{R}$ takes the values 0 and 1 with probabilities $x$ and $y$) the space $\phi_{R}$ in which $N_{R}$ acts is of rank~2.

The `augmented Hamiltonian' can then be written as :

\begin{eqnarray}
 \widetilde H&\ =\ &\sum_{R}\ \left[ \ll\epsilon\gg + \Delta\left\{(y-x) b^\dagger_{R} b_{R} + \sqrt{xy}(b^\dagger_{R}+b_{R})\right\} \right]\ a^\dagger_{R} a_{R} \ +\ldots\nonumber\\
 & & \ldots + \sum_{R}\sum_\chi \ \left[\ll t(\chi)\gg + \sqrt{xy} t^{(2)}(\chi) (b^\dagger_{R} + b_{{R}+\chi})\ + t^{(1)}(\chi)(b^\dagger_{R}+b_{{R}+\chi}) \right]\  a^\dagger_{{R}+\chi} a_{R} \end{eqnarray}
 
 Here, $a_R, a^\dagger_R$ are electron destruction and creation operators, $\Delta = (\epsilon_B-\epsilon_A)$.

The Augmented Space Theorem\cite{ast} then leads to :

\be \ll G({R},{R}',z)\gg = \langle {R} \otimes \{\emptyset\}\vert (z\tilde{I} -\widetilde{H})^{-1}\vert
{R}' \otimes \{\emptyset\} \rangle \ee 

\n The result is significant, since we have reduced the calculation of averages to one of obtaining a particular matrix element of an operator 
in the configuration space of the variable. Physically, of course, the augmented Hamiltonian is the collection of all Hamiltonians. 
 This augmented Hamiltonian is an operator in the augmented space $\Psi\eq {\cal H}\otimes \Phi$ where ${\cal H}$ is the space spanned by the 
 tight binding basis and $\Phi$ the full space of all configurations. The result is exact. Approximations can now be introduced in 
 the actual calculation of this matrix element in a controlled manner. 
 
 The same recursion method used for spatial averaging ( Haydock \etal \cite{md27}) is ideally suited for obtaining matrix elements in augmented 
 space. Since configuration averaging is an intrinsically difficult problem, we must pay the price for the above simplification. This comes in 
 the shape of the enormous rank of the augmented space. For some time it was thought that recursion on the full augmented space was not a 
 feasible proposition. However,  if randomness is homogeneous in the sense that  $p(n_R)$ is independent of the label $R$, then the full 
 augmented space has a large number of local point group and lattice translational symmetries. These have been utilized to reduce vastly the 
 rank of the effective space on which the recursion can be carried out. Recursion on augmented space can be done now with ease, even on 
 desktop computers.

\section{\bf A simple application to  disordered graphene}
It would be interesting to apply our two different techniques on a simple, yet interesting system. We have chosen disordered graphene. The choice of this example was both because
of interest in graphene and because non-ergodicity seems to be more probable in lower dimensional
systems \cite{mj}.
Disorder in graphene can significantly modify the electronic  properties because the interplay between electron-electron interactions and the 
disorder controls the low energy regime of electrons in graphene \cite{PRB06,Alam2012,Chowdhury2014,Chowdhury2015}. The 
 presence of disorder in graphene can emerge 
through various synthesis processes when there occurs an interaction with the substrates due to its exposed surface. Various kinds of nature 
of defects such as random impurities and vacancies present in graphene can change the electronic structure. Chemical 
substitution and irradiation are some of the effective methods to produce local disorder in graphene. In fact, recently, Pereira et al. 
\cite{PRB08} have shown the dramatic changes (such as localized zero modes, gap and pseudo-gap behavior and strong resonances) in the low 
energy spectrum of graphene 
within the tight binding approximation due to this local disorder. 

With this motivation, let us introduce a simple, non-trivial model of a disordered graphene sheet. This is described by a tight-binding 
Hamiltonian with a single band $p_z$ per site. Anderson \cite{pwa} studied this model in his now celebrated 1958 paper and it goes under his 
name.
\begin{equation}
 H = \sum_{R} \varepsilon_{R} \proj_{R} + \sum_{R}\sum_{R^{\prime}}  t_{RR^{\prime}} \trans_{RR^{\prime}}
\end{equation}

\n Here  $\proj_{R}$ = $\vert R\rangle\langle R\vert$ and $\trans_{\ij}$ = $\vert R\rangle\langle R^{\prime}\vert$ are projection and transfer 
operators on the space spanned by  the tight-binding  basis $\{\vert R\rangle\}$. The simplest model of a void is just the removal of an atom 
from a site $R$, putting $t_{R'R} = 0$ and $\varepsilon_R \rightarrow\ -\infty$ .
We have used the tight-binding linear muffin-tin orbitals (TB-LMTO) technique followed by a N-th order muffin-tin orbitals (NMTO) 
 downfolding to obtain the parameters of the single $p_z$ orbital per site model. The pristine graphene itself 
is a open structure and required inclusion of three empty spheres for proper space filling. When we randomly removed carbon atoms to produce 
voids,
the space had to be replaced by empty spheres. 

\begin{table}[b!]
\centering
\caption{Tight-binding parameters generated by NMTO. $\chi_n$ refer to the n-th
nearest neighbor vector on the lattice.}
\begin{tabular}{|c|cccc|}\hline
(eV) & $\varepsilon$ & t$(\chi_1)$ & t$(\chi_2)$ & t$(\chi_3)$  \\ \hline 
& -0.291   & -2.544  & 0.1668   & -0.1586\\
\hline
\end{tabular}

\label{tab1}
\end{table}

\begin{figure}
\centering
\includegraphics[width=12cm,height=18cm]{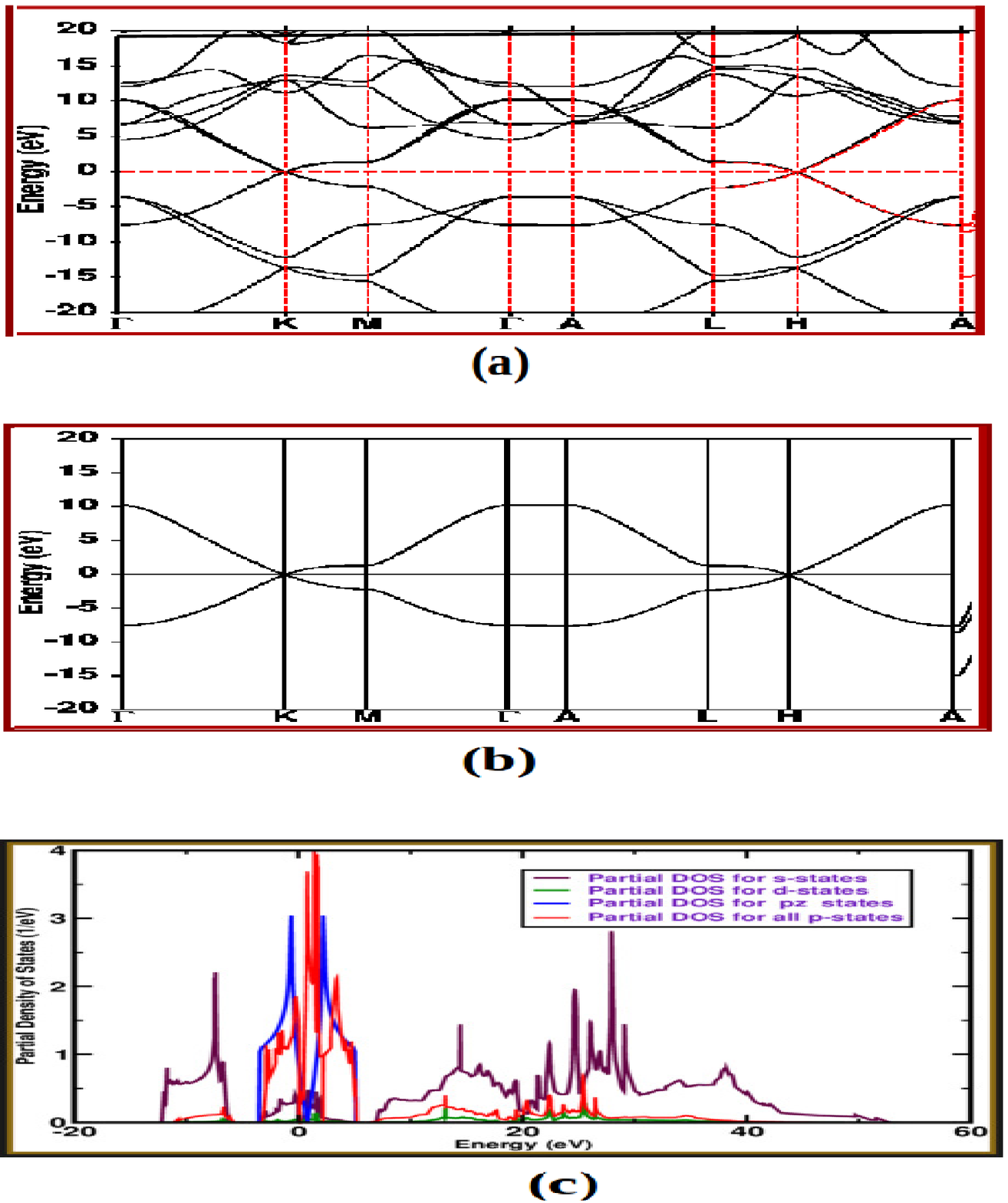}\vskip 1.5cm 
\caption{(a) The all orbital TB-LMTO band structure for graphene.(b) The downfolded  p$_z$ bands based on the single band effective Hamiltonian. 
(c) The orbital projected partial density of states.\label{fig3}}
\end{figure}

 The  tight-binding parameters are shown in Table \ref{tab1}. The overlap $t$ decays with distance. In this work we have truncated at the 
 nearest neighbor distances. The results of TB-LMTO band structure for graphene are shown in Figs.\ref{fig3}(a) and \ref{fig3}(b). For justification of 
 $p_{z}$ only downfolding, we have compared the various orbital projected partial density of states (PDOS) in Fig.\ref{fig3}(c).

\section{\bf Results and discussion}

\begin{figure}
\centering
\includegraphics[width=18cm,height=17cm]{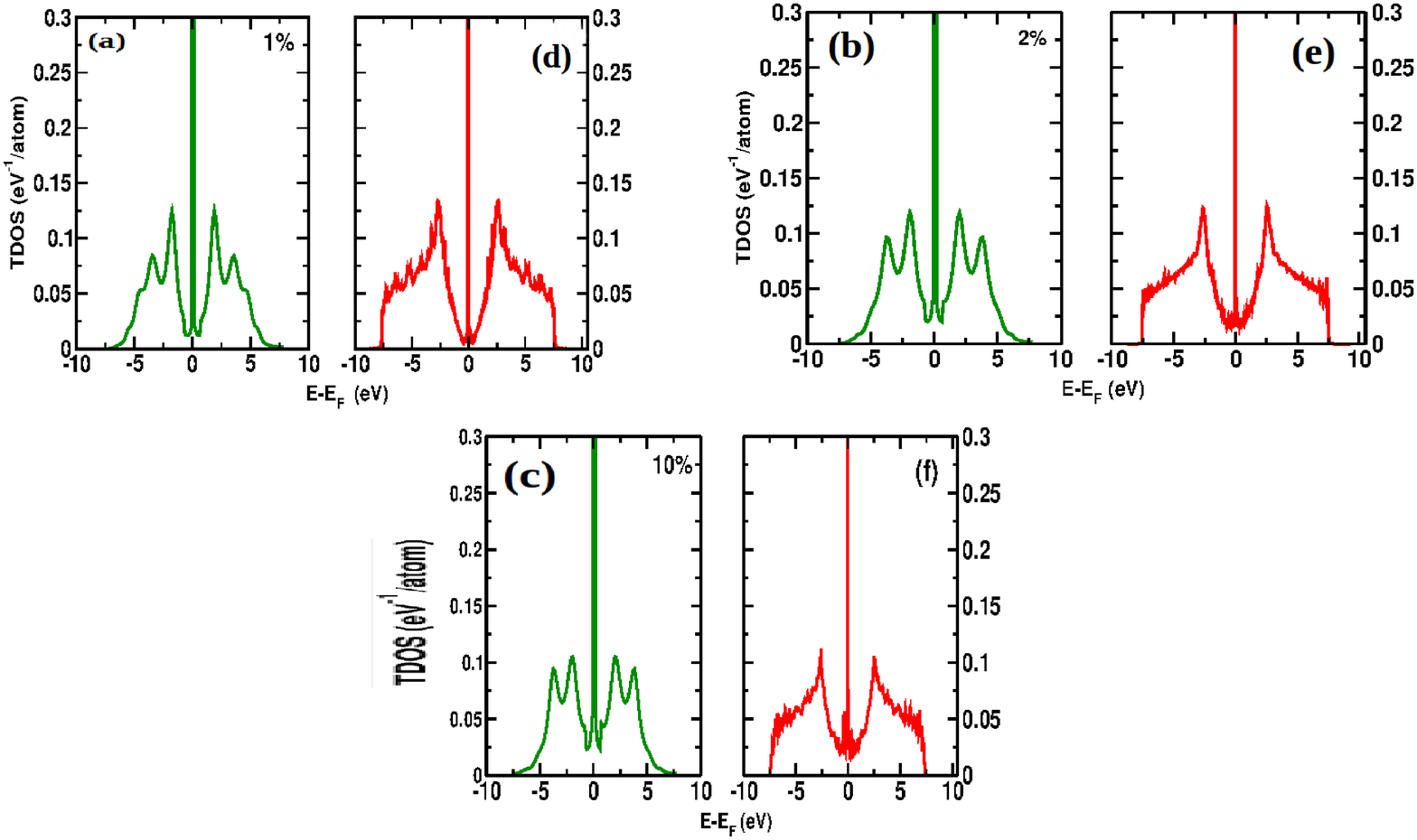}\vskip 0.1cm
\caption{(a),(b),(c):TDOS for Graphene using configuration averaging for 1, 2 and 10 \% voids respectively. (d),(e),(f):TDOS for Graphene using spatial 
averaging for 1, 2 and 10 \% voids respectively.\label{fig4}}
\end{figure}

We have applied the two above techniques : the real space recursion and augmented space to obtain the density of states (DOS) for graphene with 
random voids. The former yields the spatial averages, while the latter yields configuration averages. Figs.\ref{fig4}(a) - \ref{fig4}(f) 
compare the spatial and configuration averages of
three void compositions. Other than the energy resolution difference in the two methods (discussed shortly in the Appendix) our conclusion is 
that disordered graphene is self-averaging.

  However there is a subtle point which we should comment
 on :

 If we examine the band edges, we note that whenever there is periodicity at any scale at all, the band edges are sharp and quadratic. Long 
 range 
disorder leads to band tailing. Signs of this appear in the configuration averaged results. This disorder induced band tailing has been known 
for decades and it would be interesting to examine if the tail states are 
localized or not. 

 Again periodicity at whatever scale leads to structure in the DOS. This is smoothened out by the disorder scattering induced complex 
 `self-energy'. Occasionally practitioners of super-cell techniques introduce an artificial imaginary part to the energy. This smoothes these 
 structures too, but the procedure is entirely ad hoc. Augmented space leads to an energy dependent self-energy
systematically.

It is important to note that the inverse of the imaginary part of the self-energy, called the "life time" is accessible to neutron scattering 
experiments. The spatial averages cannot access this ``life-time" effect and it is essential to carry out configuration dependent averaging to 
calculate it. 

Further insights into different averaging procedures is desirable. Besides, the TDOS results obtained here can be verified through scanning tunneling 
microscopy and the effect on the TDOS particularly at the Fermi energy / Dirac point should reflect on the electronic transport process.

\section{Conclusions} 
We have employed two techniques, the real space recursion and augmented space to obtain the density of states for graphene with random voids. 
The numerical results obtained here clearly indicate that spatial ergodicity holds good for graphene with random voids.

\section*{Appendix}

It will be interesting to understand why different methods have different energy resolutions.	Periodicity and the Bloch Theorem leads to the 
labeling of quantum
states by a real vector $\vec{k}$ and a supplementary band label 'n'.

\[  \langle \vec{k},n\vert G(z) \vert\vec{k},n\rangle =
 \frac{1}{z-E_n(\vec{k})} \]

So that : 
\[ -(1/\pi) \sum_n Im G(E+i0,\vec{k},n) = \sum_n \delta(E-E_n(\vec{k})) \]

For a fixed $\vec{k}$ this is a set of delta functions. If we abandon periodicity and adopt the augmented space formalism :

\begin{eqnarray*}
 -(1/\pi) \sum_n Im \ll G(E+i0,\vec{k},n)\gg & = & -(1/\pi) \sum_n Im 
\frac{1}{E-E_n(\vec{k})- \Sigma_r(E,\vec{k})-i\Sigma_{im}(E,\vec{k})} \\
& = & \sum_n \frac{[1/\pi\tau(E,\vec{k},n)]}{[(E-E'_n(\vec{k})]^2 + [(1/\tau(E,\vec{k},n)]^2}
\end{eqnarray*}

This spectral density comes out to be much smoother and the smoothening lifetime
has its origin in scattering by disorder fluctuations. Moreover the lifetime
emerges out of the calculations and no external broadening or smoothening factors are required.

\section*{\bf Acknowledgments}
SC would like to thank DST, India for financial support through the Inspire Fellowship. This work was done under the HYDRA collaboration 
between our institutes. 


\begin{thebibliography}{10}
\bibitem{kn:erg1} A Cunha \textit{Physic\ae} (Arnold P. Gold Foundation (APGF), Brazil) \textbf{10} 9 (2011)
\bibitem{kn:cs} S Chandrasekhar \textit{Radiative Transfer} New York : Dover Publications Inc (1960)
\bibitem{mj} A Mookerjee and A Jayannavar \textit{Pramana} \textbf{34} 441 (1990)
\bibitem{kn:erg7} M Badino \textit{The Foundational Role of Ergodic Theory} U.S.A : Springer \textbf{11} 323 (2005)
\bibitem{kn:erg5} J van Lith \textit{Study of history and philosophy of modern physics} \textbf{32} 581 (2001)
\bibitem{ter2} R Haydock \textit{Solid State Physics : Advances in Research and Applcation} (New York : Academic Press) \textit{(eds.)} R Haydock, 
H Ehrenreich, F Seitz and D Turnbull \textbf{Vol 35,} p 215 (1980)
\bibitem{md27} R Haydock, V Heine and M Kelly \textit{J. Phys. C : Solid State Phys} \textbf{5} 2845 (1972)
\bibitem{hn}R Haydock and C Nex \textit{Phys. Rev. B} \textbf{74} 20521 (2006)
\bibitem{ter3} M Luchini and C Nex \textit{J. Phys. C : Solid State Phys}. \textbf{20} 3125 (1987)
\bibitem{bp} N Beer and D Pettifor \textit{Electronic Structure of Complex Systems} (New York : Plenum Press) \textit{(eds.)} P Phariseau and W Temmerman 
\textbf{Vol 113,} p 769 (1982)
\bibitem{ter5} V Viswanath and G M\"uller \textit{The Recursion Method, Applications to Many-Body Dynamics} (Germany : Springer-Verlag) H Araki, E Brezin, J Ehlers, 
U Frisch, K Hepp, R L Jaffe, R Kippenhahn, H A Weidenmuller, J Wess, J Zittartz and W Beiglbock (1994)
\bibitem{mag}A Magnus, D G Pettifor and D Weaire \textit{The recursion method and its applications} (New York : Springer-Verlag)  \textit{(eds.)} D G Pettifor and D Weaire 
(1984)
\bibitem{mook}A Mookerjee \textit{J. Phys. C: Solid State Phys} \textbf{6} L205 (1973)
\bibitem{tca}R Mills and P Ratanavararaksa \textit{Phys. Rev. B} \textbf{18} 5918 (1978)
\bibitem{icpa}S Ghosh, P Leath and M Cohen \textit{Phys. Rev. B} \textbf{66} 214206 (2004)
\bibitem{ast}A Mookerjee \textit{Electronic Structure of Surfaces, disordered systems and clusters} (London \& New York : Taylor \& Francis) 
A Mookerjee and D D Sarma \textbf{Vol 4,} p 168 (2003) ; (see also) T Saha Dasgupta and A Mookerjee \textit{J. Phys. Condensed Matter Phys} \textbf{6} L245 (1994)
\bibitem{PRB06} N M R Peres, F Guinea and A H Castro Neto \textit {Phys. Rev. B} \textbf {73} 125411 (2006)
\bibitem{Alam2012} A Alam, B Sanyal and A Mookerjee \textit {Phys. Rev. B} \textbf {86} 085454 (2012)
\bibitem{Chowdhury2014} S Chowdhury, S Baidya, D Nafday, S Halder, M Kabir, B Sanyal, T Saha-Dasgupta, D Jana and A Mookerjee \textit{Physica E} \textbf{61} 
191 (2014)
\bibitem{Chowdhury2015} S Chowdhury, D Jana and A Mookerjee \textit{Physica E} \textbf{74} 347 (2014)
\bibitem{PRB08} V M Pereira, J M B Lopes dos Santo and A H Castro Neto \textit {Phys. Rev. B} \textbf {77} 115109 (2008)
\bibitem{pwa} P W Anderson \textit{Phys. Rev.} \textbf{109} 1492 (1958)
\end{thebibliography}

\newpage

\begin{center}
\bf{\LARGE{Figure Captions}}
\end{center}

Figure 1 : (a)-(d) Ensemble of macrosystems in the ’configuration space’ of a binary alloy. (e) A single
macrosystem constructed out of the microsystems shown in (a)-(d).

Figure 2 : (a) Asymptotic parts of the calculated graphene Green function continued fraction coefficients obtained by recursion for $5 \%$ voids and  
(b) the same for $1 \%$ voids.

Figure 3 : (a) The all orbital TB-LMTO band structure for graphene.(b) The downfolded $p_z$ bands based on the single band effective Hamiltonian. 
(c) The orbital projected partial density of states.

Figure 4 : (a),(b),(c):TDOS for Graphene using configuration averaging for 1, 2 and 10 \% voids respectively. (d),(e),(f):TDOS for Graphene using spatial 
averaging for 1, 2 and 10 \% voids respectively.

\newpage

\newpage

\newpage

\newpage

\end{document}